\documentstyle[preprint,prb,aps]{revtex}
\begin{document}
\bibliographystyle{prsty}
\preprint{SU-ITP \# 96/11}
\title{Chiral sound waves from a gauge theory 
of 1D generalized statistics} 

\author{Silvio J. Benetton Rabello\thanks{e-mail: 
silvio@quantum.stanford.edu}}
\address{\it Department of Physics, Stanford University, Stanford CA~~94305}
\maketitle
\begin{abstract}
\end{abstract}
{\sl A topological gauge field theory in one spatial dimension is studied, 
with the gauge fields as generators of two commuting U(1) Ka\u{c}-Moody
algebras. Coupling of these gauge fields to nonrelativistic bosonic
matter fields, produces a statistical transmutation of the later, as in
the Chern-Simons theory in two dimensions. The sound waves of the model 
are investigated and proven to be chiral bosonic excitations, with the 
same spectrum as the density fluctuations of the Luttinger model.}
  

To describe a system of interacting electrons confined to move in 
one-dimensional (1D) structures one has to abandon the Landau 
Fermi liquid theory and switch to the Luttinger liquid picture
\cite{Lutt,Hald1}, due to the peculiarities of the Fermi surface in 1D. 
The Luttinger model is exactly solvable and its solution 
set the path for the representation of fermions in terms of bosonic
operators \cite{MattLieb}. For many years the Luttinger liquid remained 
as a theoretical construction with no experimental support, since it
involves a very difficult scenario to realize in practice. Recent
developments in the fabrication of nanostructures, namely in the
field of quantum wires raised the hope that such behavior could
be observed experimentally. Unfortunately due to localization by
impurities the Luttinger picture is destroyed, and no evidence
of its existence was found. In a recent experiment \cite{Mill}
 with a fractional quantum Hall  (FQH) sample, the tunneling 
conductance between two edges of a $\nu={1\over3}$ state turned
 out to show a power law dependence on the temperature that 
characterize a Luttinger liquid behavior. This is so because the
edge excitations of the FQH liquid are described by a chiral Luttinger
liquid\cite{Wen} that has an absence of localization by impurities.

Recently it was shown \cite{Wu1} that the low temperature critical 
properties of the Luttinger liquids can be encoded by an ideal gas of 
particles obeying a generalized exclusion statistics \cite{Hald2}. It is
not a great surprise that this was possible, since the Luttinger model
is solvable and in 1D the exchange of particles necessarily involves
a collision and the phase shifts due to scattering and statistics cannot
be distinguished. 

In this paper we present a variant of a previously introduced gauge 
theory of 1D generalized statistics \cite{Ra}. Similarly to the 
Chern-Simons construction for the fractional quantum Hall effect \cite{S-C},
topological gauge fields are coupled to nonrelativistic bosonic matter fields.
When the gauge sector is solved an effective Hamiltonian for the particles
emerges, this Hamiltonian contains a gauge self-interaction that can be
removed by redefinition of fields, that now obey deformed
commutation relations. The density fluctuations about a constant 
background are studied in the long wave-length limit and proven to 
be chiral bosonic excitations with the same spectrum as 
the sound waves of the Luttinger model.

Imagine a system of bosonic and spinless particles of mass m in 1D, 
coupled to some topological gauge fields. The total Hamiltonian operator 
describing this system, in the  many-body language, is given by
\begin{equation}
\label{H}
 H=\int_{-\infty}^\infty dx\biggl[ 
-{1\over 2m}\phi^\dagger(\partial_x -iA_x 
)^2\phi+A_0\phi^\dagger\phi
-A_0{1\over 2\kappa\pi}(A_x-\chi)
\biggr]\,,
\end{equation}
with $\kappa$ a real parameter, $\phi$ and $\phi^\dagger$ bosonic 
nonrelativistic matter fields, and $A_0,A_x$ and $\chi$ are the 
topological gauge fields. This system is invariant under the gauge 
transformations 
\begin{eqnarray}
\phi'(x,t)&=&\phi(x,t)e^{i\Lambda(x,t)}\,,
\qquad  \phi'^\dagger(x,t)=e^{-i\Lambda(x,t)}\phi^\dagger(x,t)\,,\\ 
A_x'(x,t)&=& A_x (x,t)+\partial_x\Lambda(x,t)\,,\qquad 
A_0'(x,t)=A_0(x,t)-\partial_t\Lambda(x,t)\,,\\
\chi'(x,t)&=&\chi(x,t)+\partial_x\Lambda(x,t)\,. 
\end{eqnarray}
The matter fields have de usual equal times commutation relations
\begin{eqnarray}
\label{matter1}
[\phi(x,t),\phi(y,t)]&=&0\,,\\
\label{matter2}
[\phi(x,t),\phi^\dagger(y,t)]&=&\delta(x-y)\,,
\end{eqnarray}
and the gauge fields generate two commuting U(1) Ka\u{c}-Moody (K-M) 
algebras
\begin{eqnarray}
\label{K-M}
[A_x(x,t),A_x(y,t)]&=&2\kappa\pi 
i\partial_x\delta(x-y)\,,\\
\label{K-M2}
[\chi(x,t),\chi(y,t)]&=&-2\kappa\pi 
i\partial_x\delta(x-y)\,,\\
\label{K-M3}
[\chi(x,t),A_x(y,t)]&=&0\,.
\end{eqnarray}

From (\ref{H}) we can extract the Gauss law operator, the generator of
time independent gauge transformations of our system, as in 
electrodynamics it is the term that multiplies $A_0$:
\begin{equation}
\label{Gausslaw}
G={1\over 2\kappa\pi}(A_x-\chi)-\phi^\dagger\phi\,.
\end{equation}
In order to solve the Gauss law, to set $G=0$ as an operator equation, 
we must pass to gauge invariant matter fields, i.e. fields $\Phi$ 
that obey $[G,\Phi]=0$. A realization of $\Phi$ is given by
\begin{equation}
\label{phys}
\Phi(x,t)=
\phi(x,t)e^{-{i\over 2}
\int_{-\infty}^x dz\,(A_x(z,t)+\chi(z,t))}\,,
\end{equation}
that together with its conjugate obeys the usual bosonic commutation 
relations (\ref{matter1},\ref{matter2}). With (\ref{H}) in terms
of $\Phi$ and setting $G=0$ we get the effective Hamiltonian
\begin{equation}
\label{ham}
H=-\int_{-\infty}^\infty dx\, 
{1\over 2m}\Phi^\dagger(\partial_x 
-i{\kappa\pi}\Phi^\dagger\Phi)^2\Phi\,.
\end{equation}
The above self-interaction can be removed with aid of a 
Jordan-Wigner transformation
\begin{equation}
\label{phys'}
{\bar\Phi}(x,t)=
\Phi(x,t)e^{-i\kappa\pi
\int_{-\infty}^x dz\,\Phi(z,t)^\dagger\Phi(z,t)}\,,
\end{equation}
so that the redefined fields obey the free Schr{\"o}dinger equation, but now
have the deformed commutation relations: 
\begin{eqnarray}
\label{anyalg'}
{\bar\Phi}(x,t){\bar\Phi}(y,t)&-&e^{i{\kappa\pi}\, 
\epsilon(x-y)}{\bar\Phi}(y,t){\bar\Phi}(x,t)=0\,,\\
{\bar\Phi}(x,t){\bar\Phi}^\dagger(y,t)&-&e^{-i{\kappa\pi }
\,\epsilon(x-y)}{\bar\Phi}^\dagger(y,t)
{\bar\Phi}(x,t)=\delta(x-y)\,,
\end{eqnarray}
with $\epsilon(x)=-\epsilon(-x)$, and the statistics is controlled by $\kappa$. 

Let us now consider the long wave-length (LWL) density fluctuations 
(sound waves) of the model. To describe these excitations start from 
(\ref{ham}), write the matter fields as $\Phi=\sqrt{\rho}e^{i\eta}$, 
and treat  $\rho$ as given by a small fluctuation about a constant background
density $\rho_0$, i.e. $\rho=\rho_0+\delta\rho$. Keeping only the terms that 
are relevant in the LWL approximation we have the Hamiltonian
\begin{equation}
\label{ham2}
H_{LWL}\approx\int_{-\infty}^\infty dx\, 
{1\over 2m}\rho_0(\partial_x\eta 
-\kappa\pi\rho)^2\,.
\end{equation}
The time evolution of the density fluctuation $\delta\rho$ is given by
\begin{equation}
\label{cboson}
\partial_t\delta\rho=i[H_{LWL},\delta\rho]\approx \kappa{\pi\rho_0\over m}
\partial_x\delta\rho\,,
\end{equation}
where again  only the long distance dominant terms were kept. 
We see that (\ref{cboson}) describes the motion of a chiral sound 
wave with velocity $v=\kappa{\pi\rho_0\over m}$. The spectrum of 
these waves can be obtained in a straightforward manner, by 
introducing the operator
\begin{equation}
\label{sound}
\sigma(x,t)=\partial_x\eta -\kappa\pi\rho\,,
\end{equation}
that in the momentum space generates a
K-M algebra
\begin{equation}
\label{sKM}
[\sigma(p,t),\sigma(-p',t)]=2\pi\kappa p\delta(p-p')
\end{equation}
From $\sigma(p,t)$ and for $\kappa>0$  
the following boson operators can be constructed:
\begin{eqnarray}
\label{Fock1}
a(p)&=&\sqrt{1\over 2\pi\kappa\vert p\vert}\sigma(p)\,,\\
a^\dagger(p)&=&\sqrt{1\over 2\pi\kappa\vert p\vert}\sigma(-p)\,,
\end{eqnarray}
for $p>0$, and
\begin{eqnarray}
\label{Fock2}
a(p)&=&\sqrt{1\over 2\pi\kappa\vert p\vert}\sigma(-p)\,,\\
a^\dagger(p)&=&\sqrt{1\over 2\pi\kappa\vert p\vert}\sigma(p)\,,
\end{eqnarray}
for $p<0$. They satisfy $[a(p),a^\dagger(p')]=\delta(p-p')$, for $\kappa<0$ 
just reverse these definitions. With these operators the 
Hamiltonian can be written as
\begin{equation}
\label{ham3}
H_{LWL}=\int_{-\infty}^\infty dp\, \varepsilon(p)a^\dagger(p)a(p) +C\,,
\end{equation}
where $C$ is a constant (infinite) and the energy 
of each sound mode is $\varepsilon(p)={\pi\rho_0\over m}\vert\kappa p\vert$,
that is the same dispersion relation found for 
the Luttinger model \cite{MattLieb}. 

This oscillator representation of $H_{LWL}$ is similar to the one used 
in solving the Luttinger model \cite{MattLieb}.
In fact, there the K-M algebra makes its appearance in the 
commutation relations of the density operators for both chiralities. 
In this case this algebra is induced by the properties of the fermionic 
vacuum. Here the density operator commutes with itself at different points, 
instead $\sigma(x,t)$ the gradient of ${\bar\Phi}$'s phase argument, is the
one that acquires a K-M algebra due to the gauge interaction. In some sense 
we have here a phase-density duality between both models.

In conclusion, it was presented here a many-body theory in 1D with a gauge
interaction, that displays a generalized statistics and has chiral 
density waves with the same linear dispersion relation as the ones found 
in the Luttinger model. 

The author is grateful to Leonid Pryadko 
and Karen Chaltikian for a stimulating 
conversation about tunneling in the FQH effect.
This work was supported by the 
CNPq (Brazilian Research Council).

\end{document}